\documentclass[12pt,a4paper]{article}
\usepackage[T1]{fontenc}
\usepackage[utf8]{inputenc}
\usepackage{mathpazo}
\usepackage{microtype}
\usepackage[a4paper,left=2.1cm,right=2.1cm,top=2.6cm,bottom=2.3cm,headheight=14pt]{geometry}
\usepackage{amsmath,amssymb}
\usepackage{booktabs,longtable,tabularx,multirow,array}
\usepackage{xcolor}
\usepackage{enumitem}
\usepackage{titlesec}
\usepackage[numbers,sort&compress]{natbib}
\usepackage{url}
\usepackage{fancyhdr}
\usepackage{lastpage}
\usepackage{caption}
\usepackage{float}
\usepackage{tikz}
\usepackage{pgfplots}
\pgfplotsset{compat=1.16}
\usetikzlibrary{arrows.meta,positioning,decorations.pathmorphing,calc,shapes.geometric}
\usepackage[colorlinks=true,linkcolor=blue!50!black,citecolor=blue!50!black,urlcolor=blue!50!black]{hyperref}

\definecolor{mdpiblue}{RGB}{0,73,144}
\setlist{nosep,leftmargin=*}

\titleformat{\section}{\normalfont\large}{\thesection.}{1.0em}{}
\titleformat{\subsection}{\normalfont}{\thesubsection.}{1.0em}{}
\titleformat{\subsubsection}{\normalfont\itshape}{\thesubsubsection.}{1.0em}{}
\titlespacing*{\section}{0pt}{16pt}{6pt}
\titlespacing*{\subsection}{0pt}{12pt}{4pt}
\titlespacing*{\subsubsection}{0pt}{12pt}{3pt}
\renewcommand{\thesubsection}{\thesection.\arabic{subsection}}

\fancypagestyle{firststyle}

\newcommand{\qeff}{q_{\rm eff}}
\newcommand{\ep}{\epsilon}
\newcommand{\Ap}{A'}

\newcommand{\ap}{\alpha'}

\begin{document}
\thispagestyle{firststyle}

\noindent{\LARGE\bfseries Millicharged Fermions as Dark Matter: Theoretical Modeling and Experimental Detection\par}
\vspace{10pt}

\noindent{ Momchil Naydenov$^{1,2}$\par
\vspace{6pt}
\noindent{\footnotesize $^{1}$\ Faculty of Physics, Sofia University, 5 J.\ Bourchier Blvd, 1164 Sofia, Bulgaria\\
$^{2}$Istituto Nazionale di Fisica Nucleare, Laboratori Nazionali di Frascati, C.P. 13, 00044 Frascati, Italy\\
 Correspondence: mnaydenov@phys.uni-sofia.bg\\[2pt]

\vspace{10pt}

\begin{center}
\begin{minipage}{0.93\textwidth}
\small
\noindent\textbf{Abstract:} Fermions carrying an effective electric charge, $\qeff\ll1$, arise when a dark sector with an unbroken (or nearly unbroken) $U(1)'$ gauge symmetry is kinetically mixed with electromagnetism, or when a fermion carries a genuinely fractional rcharge. Millicharged particles (mCPs) inherit the Pauli-exclusion constraints that govern fermionic dark matter (DM) in galaxies, while acquiring a long-range plasma phenomenology absent for neutral candidates. This review connects three strands of the literature: (i) the symmetry structure, cosmology and astrophysical bounds of millicharged fermions, with an explicit comparison of the kinetic-mixing and natural fractional charge realizations; (ii) the role of mCPs in the formation and dynamics of galactic DM halos, and the conditions under which they can be embedded in the Ruffini--Arg\"uelles--Rueda (RAR) self-gravitating core--halo model of fermionic halos; and (iii) the experimental techniques that can detect mCPs trapped on Earth, arriving from the galactic halo, or emitted by the Sun. We show that requiring a halo made of keV--MeV millicharged fermions to be collisionless imposes an upper bound on the dark fine-structure constant, $\alpha'\lesssim7\times10^{-13}$--$2\times10^{-10}$ depending on mass and velocity, which, for the Milky Way and dwarf galaxies, lies orders of magnitude below the coupling required for thermal freeze-out through $\chi\bar\chi\to A'A'$. A RAR halo built from the mCP itself therefore requires non-thermal production, a charge-symmetric
plasma and a very small dark coupling, whereas a sub-dominant mCP component can be added without
changing the halo. We close with a list of currently open theoretical and experimental challenges together with a summary of the allowed regions of the parameter space of mCPs. Throughout this work we adopt natural units where $\hbar=c=k_B=1.$\\[4pt]
\textbf{Keywords:} millicharged particles; fermionic dark matter; galactic halos; millicharged particles detection
\end{minipage}
\end{center}
\vspace{6pt}
\hrule
\vspace{4pt}

%==============================================================================
\section{Introduction}
%==============================================================================
The identity of dark matter (DM) remains one of the central open problems of contemporary fundamental physics. Among the many candidates, particles with a very small electric charge - millicharged particles (mCPs) - occupy a special position: they require no exotic interaction beyond a gauge symmetry, they arise naturally in string compactifications and in the simplest hidden-sector extensions of the Standard Model (SM), and they can be probed by electromagnetic means \cite{Holdom:1985ag,Dienes:1996zr,Abel:2008ai,Abel:2003ue,Batell:2005wa,Gherghetta:2019coi,Jaeckel:2010ni,Agrawal:2021dbo}. The simplest realization is a dark sector with a $U(1)'$ gauge symmetry whose gauge boson (the ``dark photon'' $\Ap$) mixes kinetically with the photon; a dark fermion $\chi$ that is charged only under $U(1)'$ then behaves as if it carried an electric charge $\qeff e$ with $\qeff=\ep e'/e$. Such model relaxes the need for the electric charge quantization, which is not a fundamentally emergent quality of a quantum theory.

The fermionic nature of $\chi$ has a second, largely independent, consequence. A degenerate gas of fermions cannot be squeezed into an arbitrary phase-space volume, which leads to the classic Tremaine--Gunn (TG) lower bound of order $100\,\mathrm{eV}$ on the mass of fermionic DM that is bound in dwarf galaxies solely by gravity \cite{Tremaine:1979we,Madsen:1991mz,Domcke:2014kla,Randall:2016bqw}. A parallel line of research, initiated by Ruffini and collaborators, treats galactic halos as self-gravitating, thermodynamically relaxed fermion systems (``inos'') that develop a dense degenerate core, an intermediate plateau and a classical Boltzmann tail \cite{Chavanis:2014woa,Arguelles:2014sfa,Bilic:2002kj,Arguelles:2015baa,Arguelles:2016ihf,Arguelles:2018rjj}. The Ruffini--Arg\"uelles--Rueda (RAR) model does not assume any non-gravitational interaction of the fermion. It is therefore natural to ask if the halo can be made of mCPs, what a gravitationally relevant fermion mass scale implies for the mCP experimental programme and what is the cosmological history of the inos.

The purpose of this review is to bring these questions together. The outline of the paper is as follows: Section~\ref{sec:theory} organizes the physics by its symmetry structure, distinguishes kinetic-mixing from genuinely fractional millicharges, and summarizes the effect of light charged fermions on the photon's non-local effective action. Section~\ref{sec:constraints} reviews the astrophysical, cosmological and laboratory constraints and the production mechanisms. Section~\ref{sec:halo} discusses how mCPs can influence the formation and dynamics of galactic DM halos. Section~\ref{sec:RAR} summarizes the RAR model, formulates the conditions under which an mCP can be part of it, and qualifies the argument. Sections~\ref{sec:halodet} and \ref{sec:solar} review detection of mCPs from the galactic halo and from the Sun, including a case study of the FLASH haloscope. Section~\ref{sec:challenges} collects the theoretical and experimental challenges, and Section~\ref{sec:conc} concludes. 

\subsection{Relation to existing reviews and contribution of this work}\label{sec:novelty}
Several reviews touch on parts of this subject which fall into three groups. (a) Broad surveys of particle DM, of the low-energy frontier or of feebly interacting particles, in which millicharged particles are one portal among many \cite{Bertone:2004pz,Jaeckel:2010ni,Battaglieri:2017aum,Agrawal:2021dbo}. (b) Topical reviews that cover one ingredient each: dark-photon limits \cite{Caputo:2021eaa}; searches for isolated fractional charge in matter \cite{Perl:2009zz}; small-scale structure and self-interacting DM (SIDM) \cite{Bullock:2017xww,Tulin:2017ara}; keV-scale fermions, sterile neutrinos and phase-space bounds \cite{Boyarsky:2018tvu,Drewes:2016upu}; and ultralight bosonic DM \cite{Hui:2016ltb,Antypas:2022asj}. (c) Primary papers on the recent developments most relevant here---the terrestrial thermalized mCP population, direct deflection, Cavendish tests, electrostatic accumulation, bosonic condensates, jitter-corrected Dark SRF analysis and the Tremaine--Gunn-evading Yukawa mechanism \cite{Pospelov:2020ktu,Berlin:2023zpn,Berlin:2019uco,Berlin:2023gvx,Berlin:2024ewa,Berlin:2024dwg,Berlin:2025hjs,Berlin:2025btf,Kalia:2025afc,Barir:2026wjq}---which have not, to our knowledge, been synthesized in a review. 

%==============================================================================
\section{Theoretical Background}\label{sec:theory}
%==============================================================================

Every phenomenon discussed in this review can be traced to a small number of symmetry statements. The first is abelian gauge invariance: because each abelian field strength is itself gauge invariant, the operator $F_{\mu\nu}F'^{\mu\nu}$ respects both $U(1)$ factors, and since it is renormalizable, its coefficient $\ep$ is generated by loops of any state charged under both groups and need not be small for symmetry reasons. Charge conservation under the $U(1)'$ has a further consequence, namely that the lightest charged dark state is exactly stable without any additional $Z_2$. The second statement concerns the massless limit, where the kinetic terms are invariant under $O(2)$ rotations of the vector $(A,A')$, so that only rotation invariant combinations of the couplings are physical; this explains why a massless dark photon is invisible in vacuum and becomes observable only in the presence of dark charged matter. The third is that symmetry breaking sets the range: once the dark photon acquires a mass, through either the Higgs or the Stueckelberg mechanism, the $O(2)$ redundancy is explicitly broken and the long range millicharge turns into a Yukawa interaction of range $m_{A'}^{-1}$, and it is this range that determines which experiments and which astrophysical environments are sensitive to the millicharge.

\subsection{Kinetic mixing and the effective charge}\label{sec:KM}
Consider two abelian gauge fields, the photon $A_\mu$ and the dark photon $A'_\mu$, and a dark fermion $\chi$ charged only under the latter,
\begin{equation}
\mathcal L\supset -\tfrac14 F_{\mu\nu}F^{\mu\nu}-\tfrac14 F'_{\mu\nu}F'^{\mu\nu}+\tfrac{\ep}{2}F_{\mu\nu}F'^{\mu\nu}-eA_\mu J^\mu-e'A'_\mu J'^\mu ,
\label{eq:L}
\end{equation}
with $J^\mu$ the SM current, $J'^\mu=\bar\chi\gamma^\mu\chi$ the dark current, and $e'=\sqrt{4\pi\ap}$ \cite{Holdom:1985ag,Okun:1982xi}. The kinetic terms are made canonical to leading order in $\ep$ by $A'_\mu\to A'_\mu+\ep A_\mu$, after which $\chi$ couples to the photon with strength $\qeff e$, where
\begin{equation}
\qeff=\ep\,\frac{e'}{e},
\label{eq:qeff}
\end{equation}
while SM currents couple only to $A_\mu$ \cite{Holdom:1985ag,Berlin:2023gvx,Berlin:2024ewa}. In simple field-theoretic models $\ep$ is generated at one loop by heavy particles charged under both sectors, $\ep\sim N e' e/(4\pi)^2$, which gives $\qeff\sim N\ap/(4\pi)$, i.e.\ millicharges of order $10^{-3}$ for a dark coupling comparable to $\alpha$; much smaller values are possible when $\ep$ first arises at two loops or through higher-dimension operators \cite{Abel:2008ai,Gherghetta:2019coi,Berlin:2025btf}.

In the canonical basis the charge vectors of the SM and of $\chi$ in the $(A,A')$ plane are $\mathbf Q_{\rm SM}=(e,0)$ and $\mathbf Q_\chi=(\ep e',e')$. A rotation of the plane changes the components but not lengths or scalar products, so the observable single particle couplings are
\begin{equation}
\qeff=\frac{\mathbf Q_\chi\cdot\mathbf Q_{\rm SM}}{|\mathbf Q_{\rm SM}|^2},\qquad e'_\perp=|\mathbf Q_\chi-\qeff\,\mathbf Q_{\rm SM}|\simeq e' ,
\label{eq:inv}
\end{equation}
the second being the dark charge that is felt only through the dark photon (Figure~\ref{fig:vectors}). In the absence of dark charged matter ($J'=0$) there is only one vector, $\mathbf Q_{\rm SM}$, and no invariant can be formed: the massless dark photon is unobservable. Once $J'\neq0$ there are two independent directions and Eq.~(\ref{eq:inv}) is measurable. This is precisely why light-shining-through-wall experiments (LSW), which normally lose sensitivity to a massless $\Ap$, regain it in the presence of an mCP background \cite{Berlin:2023gvx}. If $m_{A'}\neq0$ then $\chi$ behaves as millicharged only on length scales shorter than $m_{A'}^{-1}$, beyond which the interaction is Yukawa-screened \cite{Berlin:2024ewa,Berlin:2025btf}; whether $m_{A'}^{-1}$ exceeds a laboratory scale (of order metres, $m_{A'}\lesssim10^{-7}\,\mathrm{eV}$), the size of the Earth ($\sim10^{-14}$~eV) or galactic scales ($\sim10^{-27}$--$10^{-23}$~eV) determines which experiments and environments are influenced by the millicharge.

\subsection{Kinetic mixing millicharge versus genuine fractional charge}\label{sec:KMvsG}
The term ``millicharged'' is used for two structurally different situations, and much of the confusion in comparing bounds arises from not separating them. The fermion $\chi$ can carry a small charge, hence an electric charge $\qeff e$ under ordinary electromagnetism itself. The photon is the only long range gauge boson coupled to $\chi$; there is no dark photon, no dark charge $e'_\perp$, and $\chi$--$\chi$ interactions are purely electromagnetic with effective fine-structure constant $\qeff^2\alpha$ (Figure~\ref{fig:vectors}b). The theoretical cost is charge quantization and for this reason that kinetic mixing is regarded as the simplest realization of small charge when charge is quantized \cite{Berlin:2023gvx}. Searches for isolated fractionally charged particles in bulk matter target this type \cite{Perl:2009zz,Afek:2020lek}.

A second context where millicharge arises is that the fermion carries a quantized charge $e'$ (for instance unity) under its own $U(1)'$, while the millicharge $\qeff=\ep e'/e$ is an effective charge induced by mixing. Quantization is respected in each sector and $\qeff$ can take any small value. An additional light state is introduced, the dark photon of coupling $e'$ and mass $m_{A'}$, through which $\chi$ self-interacts with strength
\begin{equation}
\ap=\frac{e'^2}{4\pi}=\frac{\qeff^2\alpha}{\ep^2}\;\ge\;\qeff^2\alpha ,\qquad(\ep\le1),
\label{eq:apbound}
\end{equation}
so that the dark force between two $\chi$ is never weaker than the electromagnetic one, and is stronger by $\ep^{-2}$ when $\ep\ll1$. For $m_{A'}\to0$ the first scenario contains the second, as the limiting case $\ep=1$ (Figure~\ref{fig:vectors}).

\begin{figure}[H]
\centering
\begin{tikzpicture}[>=Stealth,scale=0.7]
\begin{scope}
 \draw[->,gray] (-0.4,0)--(4.4,0) node[right,font=\scriptsize]{$A$};
 \draw[->,gray] (0,-0.4)--(0,3.4) node[above,font=\scriptsize]{$A'$};
 \draw[->,very thick,blue!70!black] (0,0)--(3.4,0) node[below,font=\scriptsize,xshift=-4pt]{$\mathbf Q_{\rm SM}=(e,0)$};
 \draw[->,very thick,red!70!black] (0,0)--(0.8,2.9) node[above right,font=\scriptsize]{$\mathbf Q_\chi=(\ep e',e')$};
 \draw[dashed] (0.8,2.9)--(0.8,0);
 \draw[<->] (0,-0.45)--(0.8,-0.45) node[midway,below,font=\scriptsize]{$\qeff e=\ep e'$};
 \draw[->,dotted,gray!80] (0,0)--({4.0*cos(25)},{4.0*sin(25)}) node[right,font=\tiny]{$A_1$};
 \draw[->,dotted,gray!80] (0,0)--({-1.4*sin(25)},{3.0*cos(25)}) node[left,font=\tiny]{$A_2$};
 \node[font=\small] at (1.9,-1.55) {(a) Kinetic mixing};
\end{scope}
\begin{scope}[xshift=8.2cm]
 \draw[->,gray] (-0.4,0)--(4.4,0) node[right,font=\scriptsize]{$A$};
 \draw[->,gray] (0,-0.4)--(0,3.4) node[above,font=\scriptsize]{$A'$};
 \draw[->,very thick,blue!70!black] (0,0)--(3.4,0) node[above,font=\scriptsize,xshift=-6pt]{$\mathbf Q_{\rm SM}$};
 \draw[->,very thick,red!70!black] (0,0)--(1.1,0) node[below,font=\scriptsize,yshift=-2pt]{$\mathbf Q_\chi=(\qeff e,0)$};
 \node[font=\small] at (1.9,-1.55) {(b) Genuine millicharge};
\end{scope}
\end{tikzpicture}
\caption{Charge vectors in the $(A,A')$ plane (not to scale). (a) In kinetic mixing the projection of $\mathbf Q_\chi$ on $\mathbf Q_{\rm SM}$ is the observable millicharge; the orthogonal component is the dark charge. Dotted axes $A_{1,2}$ illustrate an $O(2)$ rotation, which changes components but not lengths or the scalar product. (b) A genuinely fractional charge is collinear with $\mathbf Q_{\rm SM}$.}\label{fig:vectors}
\end{figure}
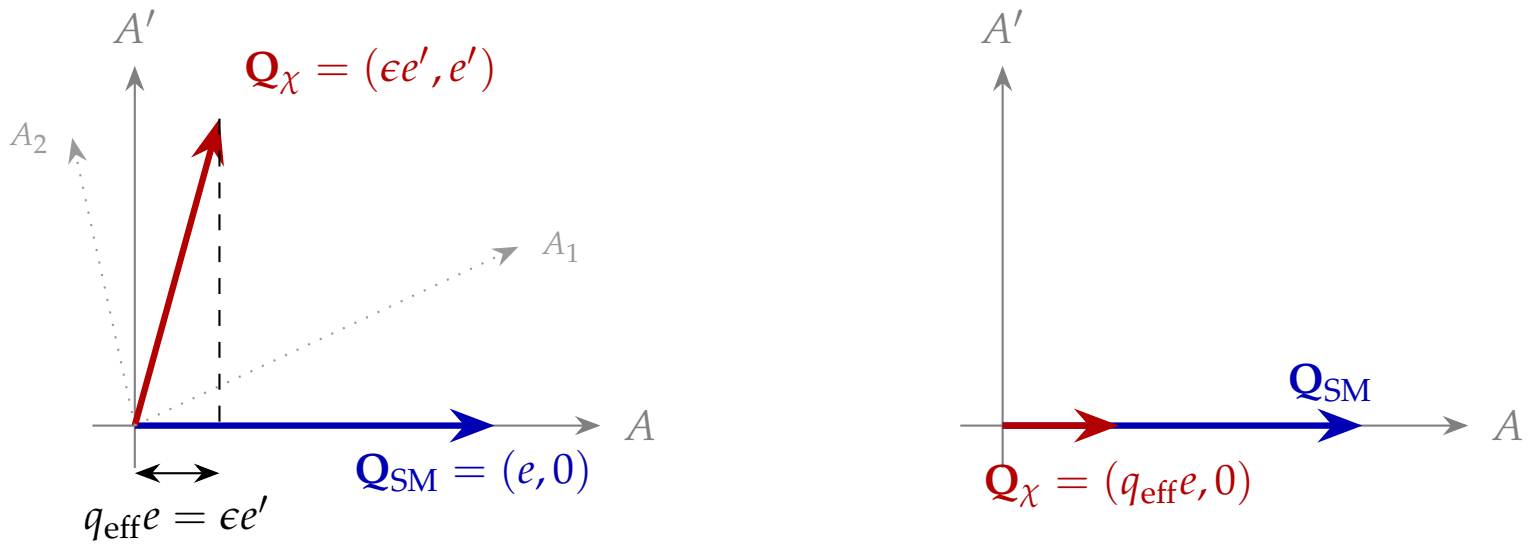

For any process involving a single $\chi$--SM vertex at momentum transfers below $m_{A'}$---stellar production by plasmon decay, accelerator production, ionization or scattering in a detector the two millicharge origins are indistinguishable at fixed $\qeff$. They differ in every observable that involves the dark sector. Two consequences are worth stating explicitly.

First, signals that rely on a dark field are specific to kinetic mixing. The LSW and direct-deflection searches of Sections~\ref{sec:halodet} and \ref{sec:solar} work because the mCP current in the emitter sources a dark electromagnetic field that passes through the shielding \cite{Berlin:2023gvx,Berlin:2024ewa}; for a purely visible charge the emitter current radiates only ordinary photons, which the walls screen (to our reading, the derivations in \cite{Berlin:2023gvx,Berlin:2024ewa} presuppose an $A'$). Conversely, static-field searches apply to both types of millicharges, but the screening is complete for the genuine millicharge type and limited to an $\ep^2$ fraction for kinetic mixing \cite{Berlin:2024dwg,Berlin:2025hjs}.

Second, Eq.~(\ref{eq:apbound}) turns a measurement of self-interactions into a test of the origin of the charge. If a millicharge $\qeff$ were measured (for instance by a Cavendish or ion-trap experiment) and a halo self-interaction strength $\ap$ inferred from clusters or dwarfs, then $\ap\simeq\qeff^2\alpha$ would indicate a genuine charge, whereas $\ap\gg\qeff^2\alpha$ would indicate mixing with $\ep\ll1$, and $\ap<\qeff^2\alpha$ would rule out a single-photon origin altogether.

\subsection{Charge conjugation and neutrality}\label{sec:C}
Two discrete symmetry choices matter for everything that follows. First, if the dark sector is $C$-symmetric, $\chi$ and $\bar\chi$ are equally abundant, the dark plasma is neutral, and the $\chi\bar\chi\to A'A'$ annihilation channel depletes the thermal abundance efficiently; the resulting relic fraction is
\begin{equation}
f_{\rm DM}\sim10^{-8}\Big(\frac{m_\chi}{1\,\mathrm{GeV}}\Big)^{2}\Big(\frac{1}{e'}\Big)^{4},
\label{eq:fDM}
\end{equation}
which is small unless $e'$ is tiny \cite{Berlin:2023gvx}. Second, if the dark sector carries a net $U(1)'$ charge, Gauss' law for the dark field forbids a homogeneous, globally charged Universe; a compensating population (e.g.\ oppositely charged heavier dark species forming dark atoms) is required \cite{Kaplan:2009de,Cyr-Racine:2012tfp}. For the discussion of galactic halos below, the symmetric (neutral) case is therefore the minimal one, and the physically relevant question is whether the resulting neutral dark plasma is collisionless on halo scales (Section~\ref{sec:RAR}).

\subsection{Running, anomalies and non-local effective actions}\label{sec:nonlocal}
Millicharged fermions lighter than the relevant momentum transfer contribute to the photon vacuum polarization in proportion to $\qeff^2$. When considering production of mCPs in curved background space-time, at one loop level we produce an effective action that is necessarily non-local

\begin{equation}
    S=-\tfrac14\!\int d^4x\,F_{\rho\sigma}\big[e^{-2}(\mu)+b\ln(\Box/\mu^2)\big]F^{\rho\sigma},
\end{equation}
with $b=\qeff^2/(12\pi^2)$ for a Dirac fermion (the corresponding scalar value is a quarter of this) \cite{Donoghue:2015xla}. Here, $e(\mu)$ is the charge running function and $\mu$ is a mass scale introduced to remove infrared divergencies. The $\ln\Box$ term is what breaks scale invariance and generates the trace anomaly $T^\mu{}_\mu=\tfrac{b e^2}{2}F_{\mu\nu}F^{\mu\nu}=\tfrac{\beta(e)}{2e}F_{\mu\nu}F^{\mu\nu}$, with $\beta=\qeff^2e^3/(12\pi^2)$; its curved-space completion additionally requires a $1/\Box$ structure related to the Riegert action \cite{Donoghue:2015xla,Donoghue:2015nba}. Covariant perturbation theory and the generalized Schwinger--DeWitt technique of Barvinsky and Vilkovisky provide the systematic framework for such non-local terms \cite{Barvinsky:1985an,Barvinsky:1990up}, complementing the local heat-kernel results of Drummond and Hathrell for a massive charged fermion in a gravitational background at high energy limit\cite{Drummond:1979pp}.

For mCPs this is mostly of conceptual interest. In the semiclassical treatment of photon deflection by a massive body, the non-local terms give a frequency-dependent correction; for the helicity-flip contribution the fractional correction to the classical bending angle scales as $\alpha\qeff^2/(3\pi)$ \cite{Donoghue:2015xla} (our rewriting of their Eq.~(102) for a fermion of charge $\qeff$), i.e.\ $\lesssim10^{-7}$ for $\qeff\lesssim10^{-2}$. Such effects are far below present sensitivity, but they illustrate a general feature: light charged particles imprint long-distance, non-analytic structure on gauge field correlators (compare the vacuum polarization contribution of a light mCP to the running of $\alpha$ and to photon dispersion in plasmas \cite{Vogel:2013raa,Masso:2006gc,DeRocco:2020xdt}), and the same non-local structure controls the collective plasma response that the detection techniques of Sections~\ref{sec:halodet} and \ref{sec:solar} exploit. We note that the non-local action papers \cite{Barvinsky:1985an,Barvinsky:1990up,Donoghue:2015xla,Donoghue:2015nba} treat massless charged fields in general and do not address dark matter physics directly; we cite them for the effective action framework and for the size of the induced gravitational effects. Additionally, such conceptual effects might turn out to be relevant if one is to calculate the mCP abundance from perturbative effects in the early Universe. The development of this concept is left for future work.

\subsection{Fermionic statistics and the Tremaine--Gunn bound}\label{sec:TG}
The Pauli principle is the statement that the many fermion state is antisymmetric under exchange, so that at most one fermion (per spin state) occupies a phase-space cell of volume $h^3$; it is the reason that a degenerate gas exerts pressure at zero temperature. For a halo of mass $M$ and radius $R$ the escape velocity scales as $v_{\rm esc}\sim\sqrt{GM/R}$ while the Fermi velocity of a degenerate gas of fermions of mass $m_\psi$ scales as $v_F\sim(M/m_\psi^4R^3)^{1/3}$; requiring $v_{\rm esc}\gtrsim v_F$ gives $m_\psi\gtrsim100\,\mathrm{eV}$ for dwarf galaxies \cite{Tremaine:1979we,Barir:2026wjq}. Degenerate fermion cores obey a Chandrasekhar-like mass--radius relation $MR^3\sim G^{-3}m_\psi^{-8}$ \cite{Domcke:2014kla,Randall:2016bqw}. The bound is a statement about the dynamics that binds the fermions: an additional attractive force replaces $G\to\alpha_s/m_\psi^2$ and relaxes it, as shown explicitly for a kpc-range scalar Yukawa force, which allows $\mathrm{eV}$ scale fermions to form dwarf-sized cores \cite{Barir:2026wjq}. Note the contrast in the symmetry of the mediator: exchange of a spin-1 boson between like charges is repulsive, so an mCP-mediated dark Coulomb force cannot provide extra binding to a charge-symmetric plasma (it averages to zero), whereas spin-0 exchange is universally attractive. This distinction is central to Section~\ref{sec:RAR}.

%==============================================================================
\section{Astrophysical, Cosmological and Laboratory Importance}\label{sec:constraints}
%==============================================================================
\subsection{Stellar and supernova bounds}
Light mCPs are produced copiously in stellar plasmas, mainly through the decay of plasmons, and contribute to loosing energy. Requiring that the Sun lose no more than about $1.5\%$ of its luminosity to mCPs (helioseismology and solar neutrinos) gives $\qeff\lesssim2\times10^{-14}$ \cite{Vinyoles:2015khy}; red-giant cooling gives $\qeff\lesssim6\times10^{-15}$ \cite{Fung:2023euv}, although the robustness of red giant branch bounds has been debated \cite{Berlin:2024ewa}. SN1987A excludes a broad region of $(m_\chi,\qeff)$ at larger charges for sub-GeV masses \cite{Chang:2018rso}, and earlier work established the general bounds \cite{Davidson:2000hf}. In-medium effects and model dependence can weaken stellar bounds \cite{Masso:2006gc,DeRocco:2020xdt,Berlin:2022hmt}: for a massive dark photon with a range shorter than the stellar size the effective charge is screened, and for $\chi$ much heavier than the core temperature production is Boltzmann suppressed---a point that is relevant for the keV--MeV masses discussed in Section~\ref{sec:RAR}.

\subsection{Accelerator and beam-dump searches}
Accelerator and beam-dump searches (SLAC, neutrino and liquid-argon experiments, LHC-based detectors such as milliQan, ArgoNeuT) exclude fractional charges from $\qeff\sim10^{-1}$ down to $\sim10^{-3}$--$10^{-5}$ for masses from MeV to tens of GeV, with future projections in the same range \cite{Battaglieri:2017aum,Prinz:1998ua,Haas:2014dda,ArgoNeuT:2019ckq,Ball:2020dnx,Magill:2018tbb,Harnik:2019zee,Plestid:2020kdm}. These searches detect relativistic $\chi$ via ionization or scintillation, and implicitly assume that the dark photon is lighter than about $\sqrt{m_eE_{\rm th}}\lesssim1\,\mathrm{MeV}$, a requirement that, combined with bounds on dark photons, implies a very light, long-ranged mediator for $\qeff\gtrsim10^{-5}$ \cite{Berlin:2025hjs}.

\subsection{Cosmology}
If $\qeff\gtrsim10^{-7}(m_\chi/\mathrm{GeV})^{1/2}$, $e^+e^-\to\chi\bar\chi$ thermalizes the mCP with the SM in the early Universe \cite{Berlin:2023gvx}. A thermalized mCP plus a massless dark photon adds dark radiation; the additional effective number of neutrino species is limited to $\Delta N_{\rm eff}\lesssim\mathcal O(0.1)$ by CMB observations \cite{Planck:2018vyg,Berlin:2024ewa,Vogel:2013raa,Berlin:2022hmt}. These limits are weakened for $m_\chi\gtrsim1\,\mathrm{GeV}$ (because the dark and visible sectors decouple before many SM mass thresholds), and can be alleviated altogether by a low reheating temperature or by non-thermal production \cite{Berlin:2023gvx}. Thermal freeze-out through $\chi\bar\chi\to A'A'$ leaves the small relic fraction of Eq.~(\ref{eq:fDM}) \cite{Berlin:2023gvx}; the complementary route is freeze-in, dominated for sub-MeV masses by the decay of in-medium plasmons and yielding a highly non-thermal (warm) phase-space distribution \cite{Hall:2009bx,Chu:2011be,Dvorkin:2019zdi}. Regarding baryon-tight coupling, CMB and 21~cm effects, before recombination, mCPs with sufficiently large $\qeff$ are tightly coupled to the baryon--photon fluid and take part in acoustic oscillations; CMB and large-scale-structure data therefore limit the fraction of DM in mCPs as a function of $(m_\chi,\qeff)$ \cite{Dubovsky:2003yn,Dolgov:2013una,McDermott:2010pa,Dvorkin:2013cea,Boddy:2018wzy,Xu:2018efh}. The 21cm absorption reported by EDGES \cite{Bowman:2018yin} and the proposal that a millicharged subcomponent could cool the baryons \cite{Munoz:2018pzp,Barkana:2018lgd} have been strongly constrained \cite{Berlin:2018sjs,Kovetz:2018zan,Stebbins:2019xjr,Creque-Sarbinowski:2019mcm,Liu:2019knx}, and an independent measurement did not confirm the EDGES profile \cite{Singh:2021mxo}. In the context of galactic magnetic fields and supernova remnants, if the dark photon is long-ranged on galactic scales, halo mCPs are magnetized. Such particles are accelerated, ejected from the disk or deflected by supernova remnants and the coherent galactic field, which gives bounds on the charge-to-mass ratio \cite{Chuzhoy:2008zy,Li:2020wyl,Stebbins:2019xjr,Dunsky:2018mqs}. These bounds do not apply if $m_{A'}^{-1}$ is shorter than the relevant magnetic coherence length, a condition that is compatible with detection on laboratory length scales. Finally, concerning cluster mergers and self-interactions, a long-ranged coherent dark force can displace dark matter relative to baryons in colliding clusters; the observed offset in the Bullet Cluster limits both the coupling and the range of such forces \cite{Markevitch:2003at,Bogorad:2023wzn}. Cosmological limits on strong dark forces of finite range are systematized in \cite{Forestell:2017wov}.

%==============================================================================
\section{Dark Matter Halo Formation}\label{sec:halo}
%==============================================================================

There have been successful theoretical attempts to obtain the mass density structure of the DM galactic halo. The leading models discuss neutral massive fermions where gravity is compensated solely by the Pauli exclusion principle which leads to a reconstruction to the most commonly used NFW profile to fit the mass distribution, avoiding the core-cusp problem. In our view, it is important to check to what extend we can link these models to specific particle theories, which are also of experimental interest. A collisionless, non-interacting fermion halo is limited by the maximum coarse-grained phase-space density. Dwarf spheroidals, which have the largest inferred coarse-grained phase-space densities among DM dominated systems, therefore set the strongest TG bounds \cite{Tremaine:1979we,Barir:2026wjq}. For fermions near the TG limit, degeneracy pressure supports cored equilibrium configurations that may even fit dwarf-galaxy data better than cuspy cold-DM profiles \cite{Domcke:2014kla,Randall:2016bqw,Bullock:2017xww}. The relaxation into such a quasi-stationary state is driven by collisionless (violent) relaxation, which for fermions naturally leads to Fermi--Dirac-like distributions \cite{Chavanis:2002yv,Chavanis:2014woa}; this is the physical motivation of the RAR framework (Section~\ref{sec:RAR}). A related effect arises in warm DM, where free streaming of light fermions erases small-scale power, so that Lyman-$\alpha$ and Milky Way satellite data bound both the mass and the phase-space distribution \cite{Bode:2000gq,Viel:2013fqw,Das:2020nwc}.

Against this background, consider first millicharged particles as a subdominant halo component. A millicharged fermion that makes up a small fraction $f_{\rm DM}$ of the halo does not alter the halo structure, but it is interesting because of how it behaves in the halo and on Earth. In galactic dynamics, if the mediator is long-ranged, halo mCPs are magnetically coupled and can be ejected from the disk or accelerated by supernova shocks, and the surviving component acts as a dark plasma with Debye screening \cite{Chuzhoy:2008zy,Stebbins:2019xjr,Li:2020wyl,Dunsky:2018mqs}. On Earth, for couplings above the thermalization threshold, galactic mCPs flowing through the planet lose their kinetic energy in the atmosphere and crust and equilibrate to $\sim300$~K; conservation of flux then implies a large enhancement of the local density, up to $\sim10^{18}$ in cryogenic detectors for $1\,\mathrm{GeV}\lesssim m_\chi\lesssim100\,\mathrm{GeV}$ \cite{Pospelov:2020ktu,Berlin:2023zpn,Berlin:2023gvx}. Terrestrial detectors with thresholds above meV are consequently insensitive, which defines the ``direct-detection ceiling'' in the $(m_\chi,\qeff)$ plane \cite{Emken:2019tni,Berlin:2023gvx}. The Sun adds a further, independent contribution: because mCPs can be produced inside it (Section~\ref{sec:solar}), the flux at Earth includes a non-galactic, non-cold component of the local mCP population.

The picture changes qualitatively if the mCP is instead the principal halo fermion, for then halo formation becomes a problem in dark plasma physics, a subject that has been explored for charged, dissipative and atomic dark matter \cite{Ackerman:2008kmp,Feng:2009mn,Agrawal:2016quu,Foot:2014uba,Fan:2013yva,Fan:2013tia,Kaplan:2009de,Cyr-Racine:2012tfp}. Three scales control the outcome. The first is Debye screening, which governs neutrality: for a charge symmetric plasma with total number density $n$ and temperature $T$, the dark Debye length is $\lambda_D=\sqrt{T/(4\pi\ap n)}$. For the local halo density $\rho\simeq0.4\,\mathrm{GeV\,cm^{-3}}$ and $m_\chi=48$~keV (so that $n\simeq8\times10^3\,\mathrm{cm^{-3}}$ and $T\simeq m_\chi v^2\simeq2\times10^{-2}$~eV for $v=200\,\mathrm{km\,s^{-1}}$), direct evaluation gives $\lambda_D\approx3\,\mathrm{km}\,(10^{-13}/\ap)^{1/2}$. For any reasonable $\ap$ this remains below a kiloparsec, so any charge separation is screened on sub-galactic scales and the plasma can be treated as neutral, which means that the mean-field equations of a neutral self-gravitating gas apply. Note, however, that this is an argument about net dark charge and does not by itself imply that collisions are negligible.

Collisionality is therefore the second scale to examine. Momentum exchange between dark charges proceeds by Rutherford scattering with a transfer cross section $\sigma_T=16\pi\ap^2\ln\Lambda/(m_\chi^2v^4)$, where $v$ is the relative velocity, the reduced mass is $m_\chi/2$, and the Coulomb logarithm is $\ln\Lambda\sim20$ (the corresponding light-mediator dynamics are discussed in \cite{Tulin:2013teo,Heikinheimo:2015kra}). Requiring $\sigma_T/m_\chi\lesssim 1\,\mathrm{cm^2\,g^{-1}}$ at the characteristic relative velocity of each system \cite{Kaplinghat:2015aga,Tulin:2017ara,Markevitch:2003at} gives, in the long-range
regime, $\alpha'_{\max}\propto v^{2}m_\chi^{3/2}$ (Appendix~\ref{app:estimates}, Eq.~\eqref{eq:amax}).
Cluster velocities ($\sim10^3$~km\,s$^{-1}$) give the weakest limit, Milky-Way-sized halos
($\sim200$~km\,s$^{-1}$) an intermediate one, and dwarf spheroidals ($\sim30$~km\,s$^{-1}$) the strongest,
a factor $(200/30)^2\simeq44$ below the Milky Way value. Figure~2 and Table~1 therefore show both the
Milky-Way and the dwarf-galaxy bounds.

The third scale is set by cooling. A charge-symmetric dark plasma radiates dark bremsstrahlung, dominated by the $\chi\bar\chi$ dipole channel, which escapes freely if the plasma is optically thin. Scaling the electron thermal bremsstrahlung emissivity by $(\ap/\alpha)^3(m_e/m_\chi)^{3/2}$, we estimate for $m_\chi=48$~keV, $\rho=0.4\,\mathrm{GeV\,cm^{-3}}$ and $T\simeq m_\chi v^2$ a cooling time $t_{\rm cool}\sim1\,\mathrm{yr}\,(\alpha/\ap)^3$. Demanding $t_{\rm cool}\gtrsim10\,\mathrm{Gyr}$ then requires $\ap\lesssim3\times10^{-6}$ ($\propto m_\chi$ at fixed $\rho$ and $v$, see Appendix); for larger $\ap$ the plasma would cool and collapse into a rotationally supported dark disk, as in dissipative dark-matter scenarios \cite{Foot:2014uba,Fan:2013yva,Fan:2013tia}. This cooling bound is weaker than the collisionality bound by many orders of magnitude and is therefore not the limiting one in Figure~\ref{fig:alpha}.

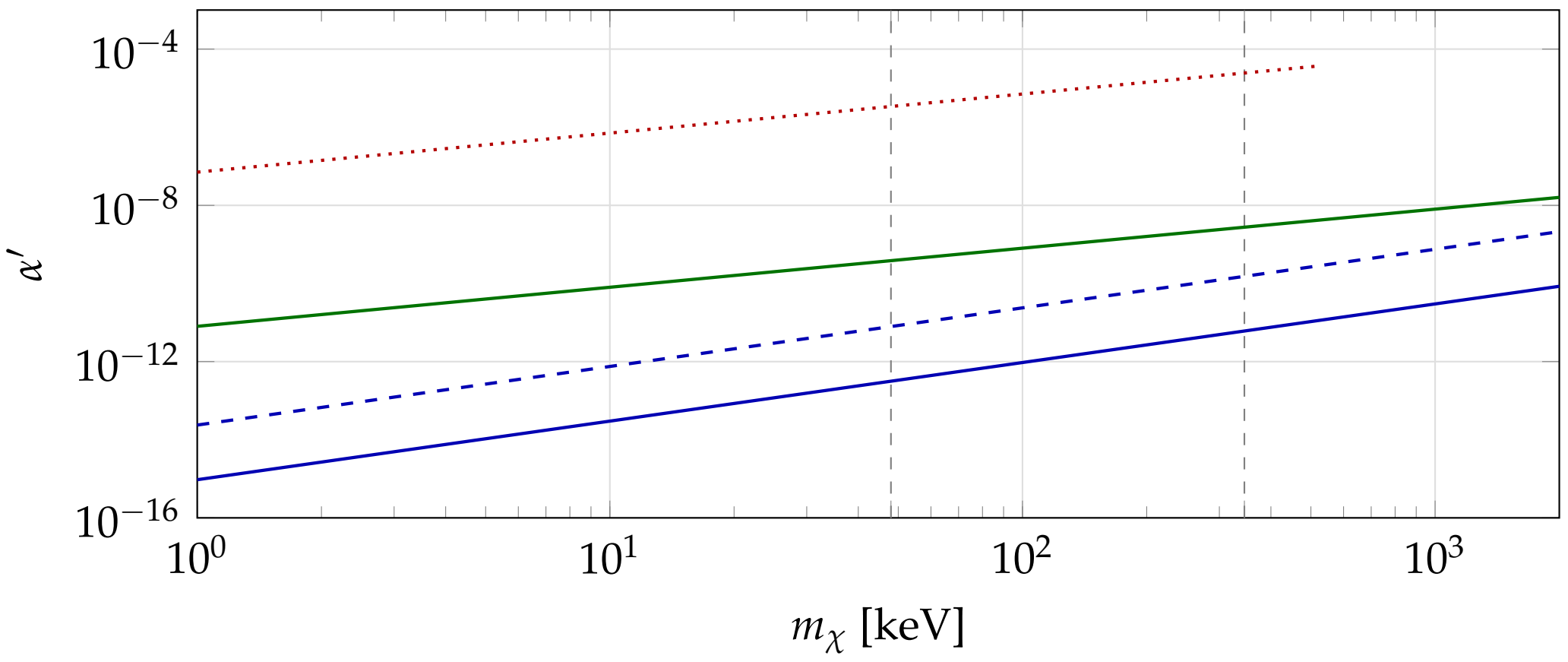
\begin{figure}[H]
\centering
\begin{tikzpicture}
\begin{loglogaxis}[
  width=0.8\textwidth, height=6.0cm,
  xlabel={$m_\chi$ [keV]}, ylabel={ $\alpha'$},
  xmin=1, xmax=2000, ymin=1e-16, ymax=1e-3,
  grid=major, grid style={gray!25}, legend pos=south east,
  legend style={font=\scriptsize, fill=white, fill opacity=0.85, draw=gray!50, at={(0.98,0.02)}, anchor=south east},
  tick label style={font=\small}, label style={font=\small},
  xtick={1,10,100,1000}, extra x ticks={48,345}, extra x tick labels={},
  extra x tick style={grid=major, grid style={dashed,gray}},
  ]
% collisionless bounds: alpha'_max = a0 (m/48 keV)^{3/2}
\addplot[thick,blue!70!black,domain=1:2000,samples=40]{3.15e-13*(x/48)^1.5};
\addplot[thick,blue!70!black,dashed,domain=1:2000,samples=40]{7.9e-12*(x/48)^1.5};
% cooling bound alpha'_max = 3.4e-6 (m/48)  (valid m<m_e)
\addplot[thick,red!70!black,dotted,domain=1:511,samples=40]{3.4e-6*(x/48)};
% thermal relic alpha'(f=1) = 8e-6 m/GeV
\addplot[thick,green!45!black,domain=1:2000,samples=40]{8e-6*(x/1e6)};

\end{loglogaxis}
\end{tikzpicture}
\caption{Constraints on the dark fine-structure constant $\ap$ for a symmetric millicharged fermion as the dominant halo component. Blue: Rutherford transfer cross section per unit mass below $1\,\mathrm{cm^2\,g^{-1}}$ (Coulomb logarithm 20). Red dotted: dark-bremsstrahlung cooling time above 10~Gyr at the local halo density. Green: $\ap$ required for $f_{\rm DM}=1$ from thermal freeze-out via $\chi\bar\chi\to A'A'$, Eq.~(\ref{eq:fDM}). Vertical dashed lines mark the RAR Milky-Way mass window $48$--$345$~keV \cite{Arguelles:2016ihf}. All curves are the authors' estimates from the cited scalings, not results of the cited papers.}\label{fig:alpha}
\end{figure}

\begin{table}[H]
\caption{Upper bounds on $\ap$ from halo collisionlessness ($\sigma_T/m_\chi<1\,\mathrm{cm^2\,g^{-1}}$, $\ln\Lambda=20$), and the maximal millicharge $\qeff=\ep e'/e$ for $\ep=1$.}\label{tab:alpha}
\footnotesize
\centering
\begin{tabular}{@{}cccc@{}}
\toprule
$m_\chi$ [keV] & $v$ [km/s] & $\ap^{\max}$ & $\qeff^{\max}(\ep{=}1)$\\
\midrule
10 & 200 & $3\times10^{-14}$ & $2\times10^{-6}$\\
10  & 30 & $7\times10^{-16}$ & $3\times10^{-7}$ \\
48 & 200 & $3\times10^{-13}$ & $6\times10^{-6}$\\
48 & 1000 & $8\times10^{-12}$ & $3\times10^{-5}$\\
48  & 30 & $7\times10^{-15}$ & $1\times10^{-6}$ \\
345 & 30 & $1\times10^{-13}$ & $4\times10^{-6}$ \\
345 & 200 & $6\times10^{-12}$ & $3\times10^{-5}$\\
345 & 1000 & $2\times10^{-10}$ & $1\times10^{-4}$\\
\bottomrule
\end{tabular}
\end{table}
Details, related to the derivation of the results in this section can be found in the Appendix.
%==============================================================================
\section{The RAR Model and Conditions for Including Millicharged Fermions}\label{sec:RAR}
%==============================================================================
\subsection{The RAR core--halo model}\label{sec:RARmodel}
The RAR model treats DM as a system of self-gravitating massive fermions ("inos") at finite temperature in general relativity \cite{Arguelles:2015baa,Arguelles:2016ihf}. With a momentum-space cutoff in the distribution function (to account for the finite halo size, following \cite{Arguelles:2014sfa}) the equation of state is
\begin{align}
\rho&=m\frac{2}{h^3}\int_0^{\epsilon_c} f_c(p)\Big[1+\frac{\epsilon(p)}{mc^2}\Big]d^3p,&
P&=\frac13\frac{2}{h^3}\int_0^{\epsilon_c} f_c(p)\,\frac{1+\epsilon/2mc^2}{1+\epsilon/mc^2}\,\epsilon\,d^3p,
\label{eq:eos}\\
f_c(\epsilon\le\epsilon_c)&=\frac{1-e^{(\epsilon-\epsilon_c)/kT}}{e^{(\epsilon-\mu)/kT}+1},&
\epsilon&=\sqrt{c^2p^2+m^2c^4}-mc^2,
\label{eq:epskin}
\end{align}
supplemented by the Einstein equations and the Tolman-Klein thermodynamic equilibrium conditions, $e^{\nu/2}T=\mathrm{const}$ and $e^{\nu/2}(\mu+mc^2)=\mathrm{const}$, and the escape (cutoff) condition $e^{\nu/2}(\epsilon+mc^2)=\mathrm{const}$ \cite{Arguelles:2016ihf}. The system is characterised by the temperature, degeneracy and cutoff parameters $\beta=kT/mc^2$, $\theta=\mu/kT$ and $W=\epsilon_c/kT$, with central values $(\beta_0,\theta_0,W_0)$ fixed, for a given fermion mass, by the halo boundary conditions. The solutions display (Figure~\ref{fig:RAR}) (i) an inner core of almost constant density governed by quantum degeneracy ($\theta>0$), (ii) an intermediate region with a sharply decreasing density followed by a plateau (where quantum corrections are still important), and (iii) a Boltzmann tail with $\rho\propto r^{-n}$, $n>2$ when the cutoff is included \cite{Arguelles:2015baa,Arguelles:2016ihf}.

\begin{figure}[H]
\centering
\begin{tikzpicture}
\begin{axis}[
  width=0.8\textwidth, height=5.0cm,
  xlabel={$\log_{10}\,r\ [\mathrm{pc}]$}, ylabel={degeneracy parameter $\theta(r)$},
  xmin=-8, xmax=5.5, ymin=-40, ymax=48, grid=both, grid style={gray!20},
  tick label style={font=\small}, label style={font=\small}, axis lines=left,
  ]
\addplot[very thick,blue!70!black,domain=-8:5.4,samples=200]{-27+64/(1+exp(2.4*(x+3.6)))};
\addplot[gray,dashed] coordinates {(-8,0) (5.5,0)};
\node[font=\scriptsize,align=center] at (axis cs:-2.5,25) {transition:\\[-2pt]$\theta:+\to-$};
\node[font=\scriptsize,align=center] at (axis cs:2.0,-12) {Boltzmann tail\\[-2pt]$\rho\propto r^{-n}$};
\node[font=\scriptsize,align=center] at (axis cs:-6.0,24) {quantum core\\[-2pt]$\theta\gg1$ (degenerate)};
\end{axis}
\end{tikzpicture}
\caption{Schematic of the three regimes of the RAR solution, drawn qualitatively after the degeneracy profile $\theta(r)$ shown for $m=48$~keV in Ref.~\cite{Arguelles:2016ihf} (central $\theta_0\simeq37$, outer $\theta\simeq-27$); the numerical location of the transition is illustrative.}\label{fig:RAR}
\end{figure}
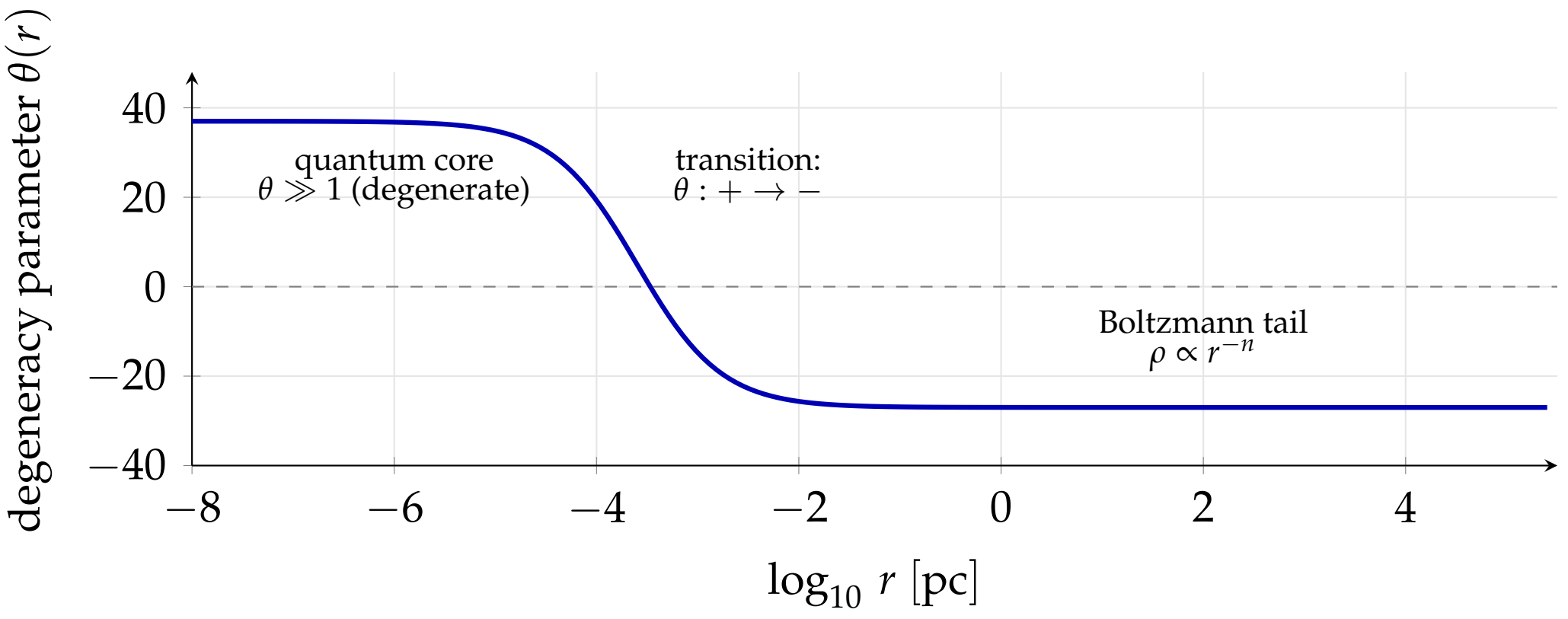

For the Milky Way the boundary conditions are the halo masses $M_{\rm DM}(40\,\mathrm{kpc})=2\times10^{11}M_\odot$ and $M_{\rm DM}(12\,\mathrm{kpc})=5\times10^{10}M_\odot$, together with a central core of $M_c=4.2\times10^{6}M_\odot$ inside the pericenter of the star S2 ($r_p=6\times10^{-4}$~pc); Ref.~\cite{Arguelles:2016ihf} finds that fermion masses $m\lesssim10$~keV are excluded because the DM rotation curve overshoots the baryon-dominated inner curve, that $10$--$48$~keV fits the halo but yields a core that is not compact enough to mimic Sgr\,A$^*$, and that $48$--$345$~keV yields a compact core, the upper limit corresponding to the last stable configuration before gravitational collapse. The extension to dwarf and larger galaxies, scaling relations and cosmological formation is developed in \cite{Arguelles:2018rjj,Arguelles:2020qsi,Krut:2023drx}, and a possible dark-matter nature of the Galactic centre compact object via the S-star orbits has been argued in \cite{Becerra-Vergara:2021gmx}; these claims must be confronted with the detection of Schwarzschild precession in the S2 orbit and with the horizon-scale image of Sgr\,A$^*$ \cite{GRAVITY:2020gka,EventHorizonTelescope:2022wkp}.

\subsection{Introducing a dark $U(1)'$ into the RAR model}
Baseline RAR fermions are neutral and interact only gravitationally (and through Pauli statistics); extensions with fermion self-interactions have been explored in \cite{Arguelles:2015baa}. A dark $U(1)'$ can enter in two logically distinct ways.

\textbf{Scenario A: mCP as a passenger.} The RAR fermion $\psi$ is dark neutral (or a sterile neutrino-like state), and a distinct fermion $\chi$ with $\qeff\neq0$ and fraction $f_{\rm DM}\ll1$ moves in the RAR gravitational potential. The RAR solution is unchanged. The mCP obeys, in the same Tolman--Klein thermodynamic equilibrium, a distribution with its own temperature and chemical potential; since $\chi$ is light and sub-dominant its density profile follows the RAR potential $e^{\nu/2}$ up to its own thermalization, and Section~\ref{sec:halodet} applies with $n^{\rm (gal)}_\chi\sim f_{\rm DM}\rho_{\rm DM}/m_\chi$. This scenario has no additional consistency conditions from the RAR side; it inherits the cosmological and stellar bounds of Section~\ref{sec:constraints}.

\textbf{Scenario B: the RAR fermion is the mCP.} Here the halo is a self-gravitating dark plasma of $\psi=\chi$. For the RAR construction to remain valid one needs the following.
\begin{enumerate}[label=(B\arabic*)]
\item \emph{Charge symmetry and neutrality.} The halo must contain equal numbers of $\chi$ and $\bar\chi$ (or another compensating species) so that the dark Poisson equation gives no net field; as shown above, $\lambda_D\ll$~kpc for any reasonable $\ap$. 
\item \emph{Collisionlessness (or a consistent collisional limit).} RAR relaxation is assumed to proceed by violent relaxation of a collisionless system; a Rutherford relaxation time comparable to the halo age would instead lead to a thermal (isothermal-core) equilibrium, which is physically different \cite{Kaplinghat:2015aga,Tulin:2017ara}. Requiring $\sigma_T/m_\chi\lesssim1\,\mathrm{cm^2\,g^{-1}}$ gives $\ap\lesssim3\times10^{-13}$ ($m_\chi=48$~keV, $v=200\,\mathrm{km\,s^{-1}}$), see Table~\ref{tab:alpha}; this is the bound at the local halo density and is only the onset of dynamically relevant collisions. For a genuine fractional charge the same bound reads $\qeff\lesssim6\times10^{-6}$, for kinetic mixing $\qeff\lesssim6\times10^{-6}\ep$.
\item \emph{Negligible dissipation.} The cooling time must exceed the halo age, $\ap\lesssim3\times10^{-6}$; automatically satisfied when (B2) holds.
\item \emph{Consistent production.} The observed DM abundance must be produced without depleting the relic by $\chi\bar\chi\to A'A'$. Thermal freeze-out would require $\ap\sim 10^{-10}$ for $m_\chi=48$~keV (Eq.~(\ref{eq:fDM}) with $f_{\rm DM}=1$), which is about three orders of magnitude above the collisionless bound of (B2) (Figure~\ref{fig:alpha}). Hence a mCP that saturates the RAR halo requires a non-thermal mechanism---plasmon-decay freeze-in \cite{Dvorkin:2019zdi,Chu:2011be}, asymmetric or late production with a very small dark coupling, $\qeff\lesssim\mathcal O(10^{-5})$--$\mathcal O(10^{-4})$ (Table~\ref{tab:alpha}). Whether this charge range survives the existing limits has to be checked case by case: for $m_\chi\sim50$--$350$~keV the solar and red-giant limits are Boltzmann-suppressed because the particle is much heavier than the core temperature, but SN1987A trapping and cooling arguments, dark-radiation limits and accelerator searches all probe this mass--charge region \cite{Chang:2018rso,Berlin:2023gvx,Vogel:2013raa}; we have not performed this evaluation, and the constraints from $\Delta N_{\rm eff}$ and the CMB depend on the production mechanism.
\item \emph{Magnetic and cluster constraints.} If $m_{A'}^{-1}$ exceeds pc-scales, magnetized mCPs are subject to disk-expulsion and supernova bounds on $q/m$ \cite{Chuzhoy:2008zy,Stebbins:2019xjr,Li:2020wyl}, which for $\qeff\lesssim10^{-5}$ and $m\sim50$~keV are difficult to evaluate without a dedicated study; a dark photon with $m_{A'}\gtrsim10^{-23}$~eV avoids them. This escape is available only for kinetic mixing type; a genuinely charged fermion always feels the galactic magnetic field.
\end{enumerate}
Conditions (B1)--(B5) can be satisfied simultaneously only in a narrow region: a charge-symmetric keV--MeV fermion with $\ap\lesssim10^{-13}$--$10^{-10}$ at the halo level, produced non-thermally, with a dark photon that is long-ranged on laboratory scales but short-ranged on galactic magnetic scales. In that regime the millicharge is dynamically irrelevant for the halo structure (the RAR solution is essentially the neutral one) but remains available as an experimental handle, e.g.\ at the level of Section~\ref{sec:halodet}: this is the sense in which mCPs can be included in the RAR framework. Another fundamental importance as to whether the RAR inos are mCPs is the production mechanism, since the cosmological history of the inos is yet unclear and is different than the hypothetical production of mCPs.

\textbf{Scenario C: attractive interactions and Tremaine--Gunn control}\label{sec:RARC}
The RAR mass window ($\sim10$--$350$~keV) is well above the gravity-only TG bound; adding an attractive force changes the picture. In Ref.~\cite{Barir:2026wjq} an ultralight scalar with a kpc range and $\alpha\sim10^{-51}$ (so that $\alpha/Gm_\psi^2\gg1$) supports degenerate cores of dwarf-galaxy size for $m_\psi=3$~eV, with a trigger sector that delays the force until $z\sim100$ and suppresses early free streaming; the force is screened on the scale of large halos. The scalar exchange is attractive between all fermions irrespective of charge, in contrast to the vector-mediated mCP force of Scenario B. A millicharged fermion that also couples to such a scalar would thus have both a gravity-like binding force and a small electromagnetic footprint; the collisionality and neutrality conditions above apply to its vector interactions, while the scalar force is governed by the constraints of \cite{Barir:2026wjq,Forestell:2017wov}. The stellar and cosmological bounds on light mCPs discussed above apply directly to the eV-scale range, where the mCP would be relativistic in stars and constrained by $\qeff\lesssim10^{-14}$--$10^{-15}$ \cite{Vinyoles:2015khy,Fung:2023euv}; whether both couplings can coexist is an open model-building question.

%==============================================================================
\section{Detection of Millicharged Particles from the Galactic Halo}\label{sec:halodet}
%==============================================================================

Strongly coupled halo mCPs thermalize in the atmosphere and crust well before reaching a detector, and their thermal energy is far below the thresholds of conventional sensors; scattering-based direct detection is insensitive above the ``direct-detection ceiling'' \cite{Emken:2019tni,Berlin:2023gvx,Iles:2024zka}. The thermalized particles nonetheless accumulate to densities that can be many orders of magnitude above the galactic average, $n_\oplus$ up to $\sim10^{15}\,\mathrm{cm^{-3}}f_{\rm DM}$ \cite{Pospelov:2020ktu,Berlin:2023zpn,Berlin:2023gvx}. Low-threshold sensors (skipper CCDs, SENSEI and DAMIC-M) are sensitive to the millicharged background in the regime below the ceiling \cite{Iles:2024zka,SENSEI:2023zdf,DAMIC-M:2025luv}.

Direct deflection and light-shining-through-wall experiments rely on detection of the collective response of the mCP fluid rather than individual scatterings \cite{Berlin:2019uco,Berlin:2021kcm}. The mCP fluid obeys, in the non-relativistic, weakly collisional limit, the momentum equation
\begin{equation}
\partial_t\mathbf V_\pm+(\mathbf V_\pm\!\cdot\!\nabla)\mathbf V_\pm+\frac{\nabla P_\pm}{n_\pm m_\chi}\simeq\pm\frac{1}{m_\chi}\big(e\qeff\mathbf E+e'\mathbf E'\big)-\Gamma_\chi(\mathbf V_\pm-\mathbf V_\mp)-\Gamma_{\rm SM}(\mathbf V_\pm-\mathbf V_{\rm SM}),
\label{eq:fluid}
\end{equation}
where $n_{\pm}$, $\mathbf V_{\pm}$ and $P_{\pm}$ are the number density, bulk velocity, and pressure of the positively or negatively charged mCP fluid, $\mathbf E'$ is the dark electric field and $\Gamma_i$ are momentum exchange rates \cite{Berlin:2023gvx}. Driving a superconducting radio-frequency (SRF) cavity in a resonant mode of frequency $\omega\sim1$~GHz and field $E_{\rm em}\sim10$~MV/m makes the mCPs oscillate; the resulting dark current $j'_{\rm em}\simeq-(i/\omega)\big[(e\qeff e' n_\chi/m_\chi)\mathbf E_{\rm em}+\omega_p'^2\mathbf E'_{\rm em}\big]$ radiates a dark electromagnetic field that penetrates the shielding and drives a visible current $\mathbf j_{\rm rec}\simeq-i(e\qeff e'n_\chi/\omega m_\chi)\mathbf E'_{\rm em}$ in a second, shielded cavity with quality factor $Q$, whose resonant response is
\begin{equation}
P_{\rm sig}\simeq\frac{Q}{\omega}\frac{\Big|\int_{\rm rec}d^3x\,\mathbf E^*_{\rm rec}\!\cdot\!\mathbf j_{\rm rec}\Big|^2}{\int_{\rm rec}d^3x\,|\mathbf E_{\rm rec}|^2}\ ,\qquad
P_{\rm sig}\propto\Big(\frac{e\,\qeff e'\,n_\chi}{m_\chi}\Big)^4 Q\,\bar E_{\rm em}^2 ,
\label{eq:Psig}
\end{equation}
with the dark plasma frequency $\omega_p'=\sqrt{e'^2n_\chi/m_\chi}$ playing the role of an effective dark-photon mass \cite{Berlin:2023gvx}. The signal disappears (exponentially suppressed) when $\omega_p'\gtrsim\omega$, and the analysis of \cite{Berlin:2023gvx} uses $\omega_p'\lesssim\omega/10$. For coaxial cavities, transverse-electric modes give the best far-field signal ($E'\propto1/z$ vs.\ $1/z^2$ for TM$_{010}$), but the penalty of the TM mode used by Dark SRF is only $\mathcal O(1)$ because the cavity separation is $d\sim1/\omega$.

Recasting the Dark SRF pathfinder run ($P_{\rm sig}<2.5\times10^{-16}$~W, $Q\simeq3\times10^{10}$, $\bar E_{\rm em}\simeq6$~MV/m) gives $\qeff\lesssim0.1\,(1/e')(m_\chi/\mathrm{GeV})(10^6\,\mathrm{cm^{-3}}/n_\chi)$ \cite{Berlin:2023gvx}. A future run ($\bar E_{\rm em}\simeq25$~MV/m, $Q=10^{12}$, thermal-noise-limited with $T=10$~mK and $t_{\rm int}=1$~yr) is projected to reach $\qeff\lesssim(1/e')(m_\chi/\mathrm{GeV})(10^6\,\mathrm{cm^{-3}}/n_\chi)\times2\times10^{-6}$ (TM$_{010}$) or $6\times10^{-7}$ (TE$_{011}$) \cite{Berlin:2023gvx}. Since $P_{\rm sig}\propto\qeff^4$ the limit scales as $P_{\rm lim}^{1/4}$. An analysis of the pathfinder data with a proper model of frequency jitter (``microphonics'') shows that the power suppression is $\sim0.87$ rather than the previously assumed $7.7\times10^{-6}$ for the dark-photon search, improving that exclusion by about an order of magnitude \cite{Kalia:2025afc,Cui:2025kxk}; the implications of this treatment for the mCP recast quoted above are not worked out in the cited works and should be revisited. The same reanalysis translates to the best laboratory limit on the photon mass, $m_\gamma\lesssim1.6\times10^{-15}$~eV \cite{Kalia:2025afc}, relevant to symmetry tests of Lorentz and gauge invariance.

\subsection{Cavendish experiments, accumulators and traps}
Cavendish tests of Coulomb's law \cite{Bartlett:1970js,Williams:1971ms} are, from the point of view of a background of mCPs, precision detectors of charges that penetrate a conducting shell: a driven outer shell at potential $\phi_0$ induces an mCP charge density $\rho_\chi\simeq\varepsilon_{\rm weak}\,m_D^2\phi_0$ inside, with Debye mass $m_D=e\qeff\sqrt{n_\chi/T_\chi}$, a photon-mass-like effect \cite{Berlin:2025hjs,Jaeckel:2009dh}. Past experiments constrain the terrestrial density down to $n_\chi\sim0.1\,\mathrm{cm^{-3}}$ and complement ion-trap limits ($n_\chi\gtrsim1\,\mathrm{cm^{-3}}$) \cite{Berlin:2025hjs,Budker:2021quh,Carney:2021irt}. An electrified ``accumulator'' shell held at fixed voltage drags positively charged mCPs inward, where they scatter and become electrically trapped, enhancing the local density by up to $\sim10^{12}(t/\mathrm{yr})$ and extending the reach of any interior sensor \cite{Berlin:2025btf}; a Cavendish experiment inside an accumulator is projected to reach the irreducible cosmic-ray-produced mCP population and to outperform future accelerator searches for sub-GeV masses \cite{Berlin:2025hjs}. Related limits on mCPs bound to matter come from levitated spheres and meteoritic searches \cite{Moore:2014yba,Afek:2020lek}. Note that in kinetically-mixed models the accumulated mCPs source a dark field that halts further accumulation once $e'^2n_\chi R_0\sim e\qeff E_0$, which limits the model-independent reach for $\ep\lesssim10^{-7}$ \cite{Berlin:2025hjs}.

\subsection{Cyclotron radiation in a magnetized haloscope: the FLASH case}\label{sec:flashcyc}
Resonant-cavity haloscopes built for axion and hidden-photon searches are natural candidates for repurposing. FLASH (Frascati, INFN-LNF) is designed around the FINUDA solenoid ($B_0$ up to $1.1$~T) and a large copper cavity ($V\simeq4.15\,\mathrm{m^3}$) covering $117$--$360$~MHz. In its nominal, magnetized configuration the only plausible channel is cyclotron emission from mCPs crossing the bore: a particle of charge $\ep e$ (here $\ep\equiv\qeff$) and mass $m$ precesses at
\begin{equation}
f_c=\frac{\ep eB_0}{2\pi m},
\label{eq:fc}
\end{equation}
so tuning the cavity to a frequency $f_c$ in the operating band fixes only the ratio, $\ep/m\in[7.4\times10^{-9},\,2.3\times10^{-8}]\,\mathrm{eV^{-1}}$ at $B_0=1.1$~T, and not the two parameters separately. Treating the trapped population as an incoherent ensemble of Larmor radiators with $n=\rho_{\rm DM}/m$ ($\rho_{\rm DM}\simeq0.4\,\mathrm{GeV\,cm^{-3}}$, $v\simeq220\,\mathrm{km\,s^{-1}}$), the power radiated along the resonance line scales linearly with the mass,
\begin{equation}
P_{\rm tot}(m)=K\,m,\qquad K=\frac{(2\pi)^4\rho_{\rm DM}V\nu_c^4v_\perp^2}{6\pi\varepsilon_0c^3B_0^2},
\label{eq:Pcyc}
\end{equation}
and demanding $P_{\rm tot}=P_{\rm sig}\simeq10^{-22}$~W (the benchmark detectable power of the FLASH design) gives threshold masses from $\sim170$~eV to $\sim14$~keV across the band, with $\ep\sim10^{-8}$--$10^{-4}$. This region of the parameter space lies inside excluded territory over its entire physically meaningful range ($m\lesssim$ few tens of MeV, beyond which $\ep>1$), by orders of magnitude.

%==============================================================================
\section{Detection of Millicharged Particles from the Sun}\label{sec:solar}
%==============================================================================
\subsection{Production and stellar bounds}

The luminocity in mCPs is $L_\chi\simeq4\times10^{-3}L_{\odot}(q_{\rm eff}/10^{-14})^2$, so that requiring $L_\chi<1.5\%L_\odot$ (helioseismology and solar neutrinos) gives $q_{\rm eff}<2\times10^{-14}.$

Analogous stellar arguments, including the dark-photon analogues, are reviewed in \cite{An:2013yfc,Redondo:2008aa}. The red-giant bound $\qeff\lesssim6\times10^{-15}$ \cite{Fung:2023euv} is less certain because it rests on modelling of the tip of the red-giant branch. Because only light mCPs ($m_\chi\lesssim$ few keV) can be thermally produced in the Sun, the solar signal is relevant to the eV--keV mass range and not to the 48--345~keV RAR window unless additional production channels exist.

\subsection{Free-streaming versus the dark solar wind}
If the dark fine-structure constant is small, solar mCPs free-stream out of the Sun with energies $\sim$keV. For larger $\ap$, number-changing processes such as $\chi\chi\to\chi\chi A'$ thermalize the mCP--$A'$ plasma inside the Sun when
\begin{equation}
\ap\gtrsim5\times10^{-6}\Big(\frac{10^{-14}}{\qeff}\Big)^{1/2},
\label{eq:therm}
\end{equation}
forming a ``dark solar wind'': a relativistic fluid, accelerated by its own thermal pressure, that is orders of magnitude denser and colder than the free-streaming population \cite{Chang:2022gcs}. At Earth, for $\qeff\sim10^{-14}$, the bulk Lorentz factor is $\gamma\simeq893$, the plasma-frame temperature and density are $\tilde T_\chi\simeq10^{-4}\,\mathrm{eV}(\qeff/10^{-14})^{1/2}$ and $\tilde n_\chi\simeq150\,\mathrm{cm^{-3}}(\qeff/10^{-14})^{3/2}$, the lab-frame energy per particle is $E_\chi\simeq0.5\,\mathrm{eV}(\qeff/10^{-14})^{1/2}$ and the plasma frequency is $\tilde\omega_p'\simeq1\,\mathrm{GHz}\,(\qeff/10^{-14})^{1/2}(\ap/10^{-5})^{1/2}$ \cite{Berlin:2024ewa}. The energy flux is about six orders of magnitude greater than the local kinetic energy flux of galactic DM \cite{Berlin:2024ewa}. Conventional elastic-scattering experiments cannot see it, since the energy exchanged in an mCP--electron collision, $\sim\mathrm{meV}\,(\qeff/10^{-14})^{1/2}$, is far below their thresholds; energetic solar mCPs may also be probed through form-factor-mediated scatterings and by boosted-DM-type analyses \cite{Chu:2018qrm,Bringmann:2018cvk,Ema:2018bih}.

\subsection{Direct deflection of solar millicharged radiation}
A driven cavity acts as a deflector for the relativistic plasma; its perturbation propagates downwind and excites a receiver cavity tuned to the same frequency. The signal current scales as $J_\chi\propto\tilde\omega_p^2$ in the weak-coupling regime, and thermalization enhances $\tilde\omega_p'$ by a factor $\sim10^3$ over the non-thermalized case \cite{Berlin:2024ewa}. The sensitivity of the pathfinder run corresponds to $|J_\chi|\simeq5\times10^{-13}\,\mathrm{A/m^2}$, while a future device ($Q=10^{12}$, $E_{\rm def}=60$~MV/m, volumes up to $\sim1\,\mathrm{m^3}$, $t_{\rm int}=1$~yr, $T_{\rm det}=10$~mK) is projected to reach $|J_\chi|\simeq3\times10^{-24}\,\mathrm{A/m^2}(1\,\mathrm{m}/L_{\rm def})^2$ \cite{Berlin:2024ewa}. The optimal window for the dark solar wind is
\begin{equation}
10^{-5}\Big(\frac{10^{-14}}{\qeff}\Big)^{1/2}\lesssim\ap\lesssim10^{-4}\Big(\frac{10^{-14}}{\qeff}\Big)\Big(\frac{1\,\mathrm{m}}{L_{\rm def}}\Big)^{2},
\label{eq:window}
\end{equation}
because for larger $\ap$ the backreaction of the plasma screens the perturbation before it reaches the detector \cite{Berlin:2024ewa}. Note that the projections assume that the axis connecting the two cavities is aligned with the Earth--Sun direction \cite{Berlin:2024ewa}. The application of this technique to a repurposed axion haloscope is discussed in Section~\ref{sec:flashdd}.

\subsection{Cosmological millicharged radiation}
The same technique also targets cosmological relativistic mCPs: with the dark plasma frequency $\tilde\omega_p'=e'\tilde T_\chi/3$ related to the energy density by $\tilde\omega_p'\simeq93\,\mathrm{GHz}\,(e'/g_*'^{1/4})(\Omega_{\rm DR}/10^{-5})^{1/4}$, a future SRF experiment can probe dark-radiation densities as small as $\sim10^{-4}$ of the CMB energy density, including millicharges of the cosmic neutrino background and dark-matter-decay or dynamical-dark-energy sources \cite{Berlin:2024ewa}. Such radiation could be present also from annihilation of halo DM into mCPs, e.g.\ $\Omega_{\rm DR}\sim10^{-4}(\mathrm{MeV}/m_{\rm DM})(\langle\sigma v\rangle/10^{-26}\,\mathrm{cm^3/s})$ from the Galactic centre \cite{Berlin:2024ewa}, providing an indirect link between halo physics and laboratory detection.

\subsection{Direct deflection at FLASH at zero magnetic field}\label{sec:flashdd}
The direct-deflection scheme \cite{Berlin:2024ewa} needs no static magnetic field. Its best sensitivity requires the extreme quality factors of superconducting (SRF) cavities ($Q\sim$ few$\times10^{10}$--$10^{12}$), but superconducting cavities tolerate ambient fields of only $\sim190$~mG, four orders of magnitude below the nominal $1.1$~T of the FINUDA magnet. An SRF-based search is therefore incompatible with FLASH in its baseline magnetized mode; the already-planned $B_0=0$ operating phase (used for FLASH's hidden-photon search) is the enabling condition \cite{Alesini:2023qed}.

Two relativistic populations are relevant to any terrestrial LSW-style detector, and both are present at FLASH's site. (i) The cold galactic halo ($\gamma\simeq1$, $v\sim10^{-3}c$): the induced current vanishes in the quasistatic limit $\omega_{\rm def}L_{\rm def}\ll1$ for $\gamma=1$ \cite{Berlin:2024ewa}, so this population is intrinsically disfavoured for direct deflection regardless of the apparatus, consistent with the cyclotron conclusion of Section~\ref{sec:flashcyc}. (ii) The dark solar wind ($\gamma\simeq893$), whose thermalization enhances $\tilde\omega_p'$ by $\sim10^3$ and, since $|J_\chi|\propto\tilde\omega_p'^2$ in the weak-coupling limit, the signal power by $\sim10^6$; this is derived from the solved fluid equations \cite{Chang:2022gcs,Berlin:2024ewa} and is the clearly favoured target.

Another detail that needs to be taken care of is the quality factor gap. The signal power scales as $P_{\rm sig}\propto Q|J_\chi|^2V_{\rm det}/\omega_{\rm def}$ \cite{Berlin:2024ewa}, so the sensitivity to $|J_\chi|$ scales as $Q^{-1/2}$. Using FLASH's copper cavity as it stands costs a factor $\sqrt{Q_{\rm SRF}/Q_{\rm Cu}}\sim10^{2}$--$10^{3.5}$ in reach relative to the Dark SRF projections, partly offset by FLASH's larger baseline volume and lower operating frequency; since the projected Dark SRF sensitivity scales as $(1\,\mathrm{m}/L_{\rm def})^2$, larger, lower-frequency cavities are favoured, which describes FLASH better than Dark SRF's compact GHz-scale cavities \cite{Berlin:2024ewa}.

The result is that the magnetized configuration offers no competitive mCP sensitivity, while the planned $B_0=0$ phase enables a qualitatively different and more promising channel whose realistic competitiveness hinges on closing the $Q$ gap with Dark SRF via an SRF insert, a two cavity shielded geometry and Earth--Sun alignment. Here we also present a summary of all excluded regions discussed in this work. 

\begin{figure}[H]
\centering
\includegraphics[width=0.75\textwidth]{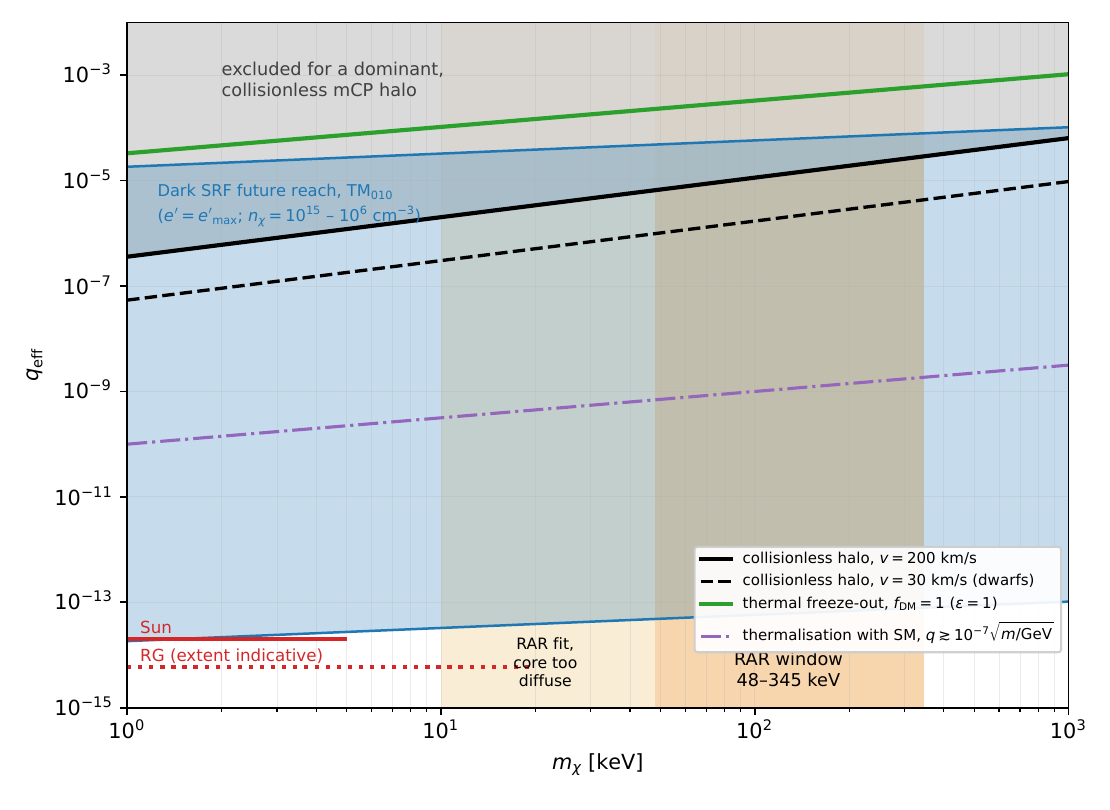}
\caption{Parameter space $(m_\chi,q_{\rm eff})$ of a millicharged fermion as the dominant halo
component ($\epsilon=1$ where $e'$ enters).
\textbf{Orange:} Milky-Way RAR mass window, $48$--$345$~keV (lighter strip: $10$--$48$~keV, core
not compact enough). It restricts $m_\chi$ only, applies only if the mCP is the RAR
fermion (Scenario~B), and is irrelevant for a sub-dominant mCP.
\textbf{Grey:} excluded for a collisionless halo, $\sigma_T/m_\chi>1\,\mathrm{cm^2\,g^{-1}}$ at
$v=200$~km\,s$^{-1}$; the solid line is $q_{\rm eff}^{\max}=\sqrt{\alpha'_{\max}/\alpha}$,
valid for any $\epsilon\le1$. The dashed line is the dwarf-galaxy bound ($v=30$~km\,s$^{-1}$).
\textbf{Blue:} projected Dark SRF reach (TM$_{010}$, $Q=10^{12}$), sensitive above the
lines, at $e'=e'_{\max}$ for $n_\chi=10^{15}$ (lower edge) to $10^{6}$~cm$^{-3}$ (upper edge); optimistic, as it
assumes Earth-thermalized enhancement.
\textbf{Green:} $q_{\rm eff}$ giving $f_{\rm DM}=1$ by thermal freeze-out, Eq.~(5).
\textbf{Purple:} thermalization threshold $10^{-7}(m_\chi/\mathrm{GeV})^{1/2}$
(extrapolated below $m_e$).
\textbf{Red:} solar and red-giant bounds; mass extents indicative.
Except for the RAR window and the stellar and thermalization thresholds, the curves are the authors'
estimates, not results of the cited papers.}
\end{figure}

%==============================================================================
\section{Open Theoretical and Experimental Challenges}\label{sec:challenges}

\subsection{Theory}
The most pressing theoretical gap is the lack of a self-consistent halo-formation calculation for a fermionic mCP with the parameters at which it could be a substantial DM component. The interplay of Pauli blocking, gravity, the vector-mediated plasma response and the possible additional scalar attraction is nowhere fully computed. A second gap is cosmological: for the keV--MeV masses most relevant to the RAR window, the constraints from $N_{\rm eff}$, CMB spectral distortions and warm-DM free streaming depend on the production mechanism, which has not been specified. A third concerns whether the RAR equilibrium is dynamically reached from cosmological initial conditions in the presence of dark-plasma effects; this has been addressed for neutral fermions in \cite{Arguelles:2020qsi} but not for charged ones.

\subsection{Experiment} 
On the experimental side the shared challenge is that signals are proportional to high powers of small couplings ($P_{\rm sig}\propto\qeff^4$) and to the local mCP density, which is model-dependent and not directly measurable; a null result constrains combinations of $n_\chi$, $\qeff$, $e'$ and $m_\chi$. Multiple complementary techniques with different density-dependence are therefore essential: a signal in one experiment must be confirmed by others, for instance by correlating with accumulator voltage, run-time and size \cite{Berlin:2025hjs}. For solar mCPs, aligning experiment geometry with the Earth--Sun axis and mitigating the effect of solar and terrestrial magnetic fields on long-ranged mCP radiation are open engineering issues \cite{Berlin:2024ewa}.

%==============================================================================
\section{Conclusion}\label{sec:conc}
%==============================================================================
Millicharged fermions sit at the intersection of gauge symmetry, statistical constraints on fermionic phase space, halo dynamics, and precision electromagnetic metrology. The symmetry structure of $U(1)\times U(1)'$ explains why massless dark photons become observable only in the presence of mCPs, through the $O(2)$-invariant charge $\qeff$ and the orthogonal dark charge, and provides a parametrically clear map between the dark-photon range and the environments in which an mCP behaves as an electrically charged particle. Kinetic mixing induced and genuinely fractional millicharges are indistinguishable in single-vertex processes, but differ in their self-interaction strength, with $\ap\geq\qeff^2\alpha$ in the kinetic mixing case, as well as in their dark radiation phenomenology and in the applicability of dark field based searches such as LSW and direct deflection, which are available only for kinetic mixing. The relation $\ap=\qeff^2\alpha/\ep^2$ therefore provides a possible route to distinguishing the two scenarios.

As a subdominant halo component, mCPs can be accommodated within a RAR-type core--halo without modifying its structure. Their experimental signatures instead arise from thermalization on Earth and can be probed through direct-deflection, LSW, Cavendish, accumulator, and ion-trap techniques. If, instead, the mCP constitutes the dominant halo fermion, it can be incorporated into the RAR construction only within a narrow parameter window: the population must be charge-symmetric and non-thermally produced, while the dark fine structure constant must be sufficiently small, $\ap\lesssim10^{-13}$--$10^{-10}$ according to our halo level collisionality estimate, for the millicharge to remain dynamically irrelevant on halo scales. Requiring the degenerate core to remain collisionless imposes an even stronger condition, $\qeff\lesssim10^{-9}$--$10^{-8}$ in our estimate; otherwise, the core becomes collisionally thermalized and its subsequent gravothermal evolution must be investigated. A two-species RAR solution, for which the zero-temperature scaling shifts the window by $2^{-1/4}$, together with a dedicated dark-plasma analysis, is needed to establish these conclusions quantitatively. More generally, the RAR window is determined by core compactness and collapse rather than by the Tremaine--Gunn bound, while its Milky-Way realization is a boundary-value construction whose uncertainty has not yet been quantified.

Tremaine--Gunn-evading mechanisms based on an attractive scalar force and RAR-type self-gravitating fermions therefore provide complementary routes to light fermionic dark matter. A model combining either mechanism with a millicharge, however, remains to be constructed. In addition to terrestrial signatures, solar mCPs, particularly in the thermalized dark-solar-wind regime, and cosmological millicharged radiation provide further gravity-independent targets for the same SRF instruments. The FLASH case study indicates that a magnetized haloscope is not competitive through its cyclotron channel, which is excluded by many orders of magnitude, whereas its planned $B_0=0$ phase could enable direct deflection of the dark solar wind if the quality-factor gap relative to SRF cavities can be closed.

Further progress will require dedicated simulations, detailed cosmological histories for specific production mechanisms, and next-generation SRF and accumulator experiments. These developments should provide quantitative answers to whether the mCP parameter window and the fermionic-halo window can overlap.

\section{Appendix}

\subsection{Derivation of the halo level estimates of Section~4}
\label{app:estimates}

For benchmark values we choose $m_\chi=48$~keV, $\rho=0.4\,\mathrm{GeV\,cm^{-3}}$, hence
$n=\rho/m_\chi\simeq8.3\times10^{3}\,\mathrm{cm^{-3}}$, and $v=200$~km\,s$^{-1}$, so
$T\simeq m_\chi v^2\simeq2.1\times10^{-2}$~eV. The plasma is charge symmetric
($n_\pm=n/2$), and $\alpha'=e'^2/4\pi$. For these values we give details on the estimates of the Debye wavelength, momentum transfer cross section and the dark bremmstrahlung cooling. 

For a two component plasma, the Debye length is $\lambda_D^{-2}=\sum_i 4\pi\alpha' n_i/T=4\pi\alpha' n/T$, so
\begin{equation}
\lambda_D=\sqrt{\frac{T}{4\pi\alpha' n}}\;.
\end{equation}
With $n=6.4\times10^{-11}\,\mathrm{eV^3}$ this gives
$\lambda_D^2=2.7\times10^{7}\,\alpha'^{-1}\,\mathrm{eV^{-2}}$, i.e.
$\lambda_D\simeq3\,\mathrm{km}\,(10^{-13}/\alpha')^{1/2}$.

\subsection{Momentum transfer cross section}
For two particles of reduced mass $\mu=m_\chi/2$ and relative velocity $v$ interacting through a
screened potential $V=\pm\alpha' e^{-m_{A'}r}/r$, the Born amplitude gives, with $q=2\mu v\sin(\theta/2)$,
\begin{equation}
\frac{d\sigma}{d\Omega}=\frac{(2\mu\alpha')^2}{(q^2+m_{A'}^2)^2}.
\end{equation}
Weighting by $(1-\cos\theta)=q^2/2\mu^2v^2$ and using $d\Omega=\pi\,dq^2/\mu^2v^2$,
\begin{equation}
\sigma_T=\frac{2\pi\alpha'^2}{\mu^2v^4}\int_0^{Q^2}\!\frac{q^2\,dq^2}{(q^2+m_{A'}^2)^2}
=\frac{8\pi\alpha'^2}{m_\chi^2v^4}\left[\ln(1+b)-\frac{b}{1+b}\right],
\qquad b=\frac{m_\chi^2v^2}{m_{A'}^2},
\label{eq:sigmaT}
\end{equation}
with $Q=2\mu v=m_\chi v$. The result is the same for $\chi\chi$ and $\chi\bar\chi$ at Born level.
For $b\gg1$ (long-range regime), we get
\begin{equation}
\sigma_T=\frac{16\pi\alpha'^2\ln\Lambda}{m_\chi^2v^4},
\end{equation}
which is the expression used in the main text. For a massless mediator, $m_{A'}\to\lambda_D^{-1}$
and $\ln\Lambda=\ln(m_\chi v\lambda_D)\simeq27$ at $\alpha'=3\times10^{-13}$; the adopted value
$\ln\Lambda=20$ changes $\alpha'_{\max}$ by only $\sim15\%$. The Born approximation is valid since
$\alpha'/v\sim10^{-9}\ll1$. For $b\ll1$ (short-range regime), $\sigma_T\to4\pi\alpha'^2m_\chi^2/m_{A'}^4$,
which is independent of $v$.

Imposing $\sigma_T/m_\chi\le(\sigma/m)_{\max}$ in the long-range regime gives
\begin{equation}
\alpha'_{\max}=\frac{v^2m_\chi^{3/2}}{4}\sqrt{\frac{(\sigma/m)_{\max}}{\pi\ln\Lambda}}
\simeq3.2\times10^{-13}\left(\frac{m_\chi}{48\,\mathrm{keV}}\right)^{3/2}
\left(\frac{v}{200\,\mathrm{km/s}}\right)^{2}
\left(\frac{20}{\ln\Lambda}\right)^{1/2}
\left[\frac{(\sigma/m)_{\max}}{1\,\mathrm{cm^2g^{-1}}}\right]^{1/2}.
\label{eq:amax}
\end{equation}
This reproduces Table~1. For a genuine fractional charge, or kinetic mixing with $\epsilon=1$,
$q_{\rm eff}^{\max}=\sqrt{\alpha'_{\max}/\alpha}$; for kinetic mixing in general, $q_{\rm eff}\le\epsilon\sqrt{\alpha'_{\max}/\alpha}$.

\subsection{Dark-bremsstrahlung cooling}
The electron--ion thermal bremsstrahlung emissivity is
$\varepsilon_{e}\simeq1.4\times10^{-27}\,T_{\rm K}^{1/2}\,n_en_i\ \mathrm{erg\,cm^{-3}s^{-1}}$.
Since the power scales as $\alpha^3m^{-3/2}T^{1/2}n_+n_-$, for the $\chi\bar\chi$ plasma
\begin{equation}
\varepsilon_\chi\simeq\varepsilon_e\!\left(T,\,n_+n_-=\tfrac{n^2}{4}\right)
\left(\frac{\alpha'}{\alpha}\right)^{3}\left(\frac{m_e}{m_\chi}\right)^{3/2}.
\end{equation}
For the benchmark, $T=247$~K gives $\varepsilon_e=3.8\times10^{-19}$ and, with $(m_e/m_\chi)^{3/2}=34.7$,
$\varepsilon_\chi=1.3\times10^{-17}(\alpha'/\alpha)^3\,\mathrm{erg\,cm^{-3}s^{-1}}$. With thermal energy density
$u=\tfrac32nT=4.3\times10^{-10}\,\mathrm{erg\,cm^{-3}}$,
\begin{equation}
t_{\rm cool}=\frac{u}{\varepsilon_\chi}\simeq3\times10^{7}\,\mathrm{s}\left(\frac{\alpha}{\alpha'}\right)^{3}
\simeq1\,\mathrm{yr}\left(\frac{\alpha}{\alpha'}\right)^{3}.
\end{equation}
Requiring $t_{\rm cool}\ge10$~Gyr gives $\alpha'\lesssim3\times10^{-6}$. At fixed $\rho$ and $v$,
$\varepsilon\propto\alpha'^3\rho^2v\,m_\chi^{-3}$ and $u\propto\rho v^2$, so $\alpha'_{\rm cool}\propto m_\chi$.
This is an order-of-magnitude estimate: $O(1)$ factors from the dipole structure of $\chi\bar\chi$
scattering are not included. It is about seven orders of magnitude weaker than Eq.~\eqref{eq:amax}.
\bibliographystyle{apsrev4-1}
\bibliography{references}

%\begingroup
%\footnotesize
%\renewcommand{\refname}{References}
%\begin{thebibliography}{999}
%\setlength{\itemsep}{0.5pt}\setlength{\parsep}{0pt}\setlength{\parskip}{0pt}
%\raggedright

\end{document}